\documentstyle[aps,prl,twocolumn,epsf]{revtex}
\begin{document}
\draft
\wideabs{
\title{Bose-Einstein condensation in quasi2D trapped gases}
\author{D.S. Petrov${}^{1,2}$, M. Holzmann${}^3$, 
and G.V. Shlyapnikov${}^{1,2,4}$}
\address{${}^1$ FOM Institute for Atomic and Molecular Physics, Kruislaan 407, 
1098 SJ Amsterdam, The Netherlands \\
${}^2$ Russian Research Center, Kurchatov Institute, 
Kurchatov Square, 123182 Moscow, Russia \\
${}^3\!$ CNRS-Laboratoire de Physique Statistique
and ${}^4$ Laboratoire Kastler Brossel $^{*}$,\\
Ecole Normale Sup{\'{e}}rieure, 24, Rue Lhomond, 
F-75231 Paris Cedex 05, France}

\date{\today}
\maketitle
\begin{abstract}
We discuss BEC in (quasi)2D trapped gases and find that well 
below the transition temperature $T_c$ the equilibrium state is a true  
condensate, whereas at intermediate temperatures $T<T_c$ one has a
quasicondensate (condensate with fluctuating phase).
The mean-field interaction in a quasi2D gas is sensitive to 
the frequency $\omega_0$ of the (tight) confinement in the "frozen" direction, 
and one can switch the sign of the interaction by changing $\omega_0$. 
Variation of $\omega_0$ can also reduce the rates of inelastic processes. 
This offers promising prospects for tunable BEC in
trapped quasi2D gases. 
\end{abstract}
}
\narrowtext

The influence of dimensionality of the system of bosons on the presence 
and character of Bose-Einstein condensation (BEC) and superfluid phase 
transition has been a subject of extensive studies in spatially 
homogeneous systems. In 2D a true condensate can only exist
at $T=0$, and its absence at finite temperatures follows from the
Bogolyubov $k^{-2}$ theorem and originates from long-wave fluctuations
of the phase (see, e.g., \cite{LL,Popov}). However, as was first
pointed out by Kane and Kadanoff \cite{KK} and then proved by Berezinskii
\cite{Ber}, there is a superfluid phase transition at sufficiently low $T$. 
Kosterlitz and Thouless \cite{KTT} found that this transition is associated 
with the formation of bound 
pairs of vortices below the critical temperature $T_{KT}=(\pi\hbar^2/2m)n_s$
($m$ is the atom mass, and $n_s$ the superfluid density just below $T_{KT}$).
Earlier 
theoretical studies of 2D systems have been reviewed in \cite{Popov} and 
have led to the conclusion that below the Kosterlitz-Thouless 
Transition (KTT) temperature the Bose liquid (gas) is characterized by the presence 
of a quasicondensate, that is a condensate with fluctuating phase (see
\cite{Kagan87}). In this case the system can be divided into blocks with a 
characteristic size greatly exceeding the healing length but smaller
than the radius of phase fluctuations. Then, there is a true condensate
in each block but the phases of different blocks are not correlated 
with each other. 
  
The KTT has been observed in monolayers of liquid helium 
\cite{Reppy}.
The only dilute atomic system studied thus far
was a 2D gas of spin-polarized atomic hydrogen on 
liquid-helium surface (see \cite{Jook} for review). 
Recently, the observation of KTT in this system has been
reported \cite{Simo}.

BEC in trapped 2D gases is expected to be qualitatively different.
The trapping potential introduces a finite size of the  
sample, which sets a lower bound for the momentum of excitations 
and reduces the phase fluctuations. 
Moreover, for an ideal 2D Bose gas in a harmonic potential 
Bagnato and Kleppner \cite{Kleppner} found a macroscopic occupation of 
the ground state of the trap (ordinary BEC) at 
temperatures $T\!\!<\!T_c\!\approx\!N^{1/2}\hbar\omega$, 
where $N$ is the number of particles, and $\omega$ the trap frequency.   
Thus, there is a question of whether an interacting trapped 2D gas supports
the ordinary BEC or the KTT type of a cross-over to the BEC regime \cite{Shev}.
However, the critical temperature will be always comparable with 
$T_c$ of an ideal gas:
On approaching $T_c$ from above, the gas density is
$n_c \sim N/R_{T_c}^2$, where $R_{T_c}\approx\sqrt{T_c/m\omega^2}$ is the thermal 
size of the cloud, and hence the KTT temperature is
$\sim \hbar^2 n_c/m \sim N^{1/2} \hbar \omega\approx T_c$.

The discovery of BEC in trapped alkali-atom clouds 
\cite{discovery} stimulated a progress in optical cooling and trapping
of atoms.
Present facilities allow one to tightly confine the motion of trapped 
particles in one direction and to create a (quasi)2D gas. In other words, 
kinematically the gas is 2D, and in the "frozen" direction the 
particles undergo zero point oscillations. This requires the frequency 
of the tight confinement $\omega_0$ to be much larger than the gas temperature 
$T$ and the mean-field interparticle interaction $n_0g$ ($n_0$ is the 
gas density, and $g$ the coupling constant). Recent experiments 
\cite{Mlynek,Chu,Christ} indicate a realistic possibility of creating 
quasi2D trapped gases and achieving the regime of quantum degeneracy 
in these systems. The character of BEC will be similar to that in 
purely 2D trapped gases, and the main difference is related to the 
sign and value of the coupling constant $g$. 

In this Letter we discuss BEC in quasi2D trapped gases and arrive at two
key conclusions. First, well below $T_c$ the phase fluctuations are
small, and the equilibrium state is a true  
condensate. At intermediate temperatures $T<T_c$ the phase fluctuates on a
distance scale smaller than the Thomas-Fermi size of the gas, 
and one has a quasicondensate (condensate with fluctuating phase).
Second, in quasi2D the coupling constant $g$ is sensitive to the frequency of
the tight confinement $\omega_0$ and, for a negative 3D scattering length $a$, one
can switch the mean-field interaction from attractive to repulsive
by increasing $\omega_0$. 
Variation of $\omega_0$ can also reduce the rates of inelastic processes. 
These findings are promising for tunable BEC.

In a weakly interacting Bose-condensed gas the correlation 
(healing) length $\hbar/\sqrt{mn_0g}$ ($g>0$) should greatly exceed the mean 
interparticle separation. In (quasi)2D the latter is $\sim 1/\sqrt{2\pi n_0}$, 
and we obtain a 
small parameter of the theory, $(mg/2\pi \hbar^2)\ll 1$ (see \cite{Kagan87}). 

We first analyze the character of BEC in a harmonically
trapped 2D gas with repulsive interparticle interaction, relying on the
calculation of the one-particle density matrix.
Similarly to the spatially homogeneous case \cite{LL,Popov}, at
sufficiently low temperatures only
phase fluctuations are relevant. Then the field operator
can be written as $\hat\Psi(\!{\bf R}\!)\!\!=\!\!n_0^{\!1\!/\!2}\!(\!{\bf R}\!)
\!\exp\{\!i\hat\phi(\!{\bf R}\!)\!\}$,
where $\hat\phi(\!{\bf R}\!)$ is the operator of the phase fluctuations,
and $n_0(\!{\bf R}\!)$ the condensate density at $T\!\!=\!\!0$. 
The one-particle density matrix takes the form \cite{Popov} 
\begin{equation}     \label{matrix}
\langle\hat\Psi^{\dagger}({\bf R})\hat\Psi(0)\rangle=
\sqrt{n_0({\bf R})n_0(0)}\exp\{-\langle(\delta\hat\phi({\bf R}))^2
\rangle/2\}.
\end{equation}
Here $\delta\hat\phi({\bf R})=\hat\phi({\bf R})-\hat\phi(0)$, and
${\bf R}=0$ at the trap center.
For a trapped gas the operator $\hat\phi({\bf R})$ is given by 
\begin{equation}     \label{pf}
\hat\phi({\bf R})=\sum_{\nu} [4 n_0({\bf R})]^{-1/2}
f_{\nu}^{+} \hat{a}_{\nu} \, + \, \mbox{h.c.},
\end{equation}
were $ \hat{a}_{\nu}$ is the annihilation operator of an 
elementary excitation with energy $\epsilon_{\nu}$, and
$f_{\nu}^{\pm}= u_{\nu} \pm v_{\nu}$ are the Bogolyubov 
$u,v$ functions of the excitations.

In the Thomas-Fermi regime the density $n_0({\bf R})$  has the well-known 
parabolic shape, with the
maximum value $n_{0m}=n_0(0)\approx(Nm/\pi g)^{1/2}\omega$, and the  
radius $R_{TF}\approx (2\mu/m\omega^2)^{1/2}$. The chemical
potential is $\mu=n_{0m}g$, and the ratio 
$T_c/\mu\approx (\pi\hbar^2/mg)^{1/2}\gg 1$. 
For calculating the mean square fluctuations of the phase, we explicitly found
the (discrete) spectrum and wavefunctions of excitations with
energies $\varepsilon_{\nu}\ll\mu$ by using the method 
developed for 3D trapped condensates \cite{exc}.  For excitations with 
higher energies we used the WKB approach.
At distances $R$ greatly exceeding the wavelength
$\lambdabar_T$ of thermal excitations ($\varepsilon_{\nu}\!\approx\!T$) near
the trap center, for $T\!\gg\!\mu$ we obtain
\begin{equation}     \label{ff}
\langle (\delta\hat 
\phi({\bf R}))^2\rangle\approx\frac{mT}{\pi\hbar^2n_{0m}}
\log{(R/\lambdabar_T)}.
\end{equation} 
We also find that Eq.(\ref{ff}) holds at any $T$ for a homogeneous gas of
density $n_{0m}$, where at $T\ll\mu$ it reproduces the well-known result
(see \cite{Popov}). In a trapped gas for $T\ll\mu$, due to the contribution of 
low-energy excitations, Eq.(\ref{ff}) acquires a numerical coefficient 
ranging from 1 at $R\ll R_{TF}$ to approximately 3 at $R\approx R_{TF}$. 

The character of the Bose-condensed state is determined by the 
phase fluctuations at $R\sim R_{TF}$. With 
logarithmic accuracy, from Eq.(\ref{ff}) we find
\begin{equation}
\langle (\delta \phi(R_{TF}))^2\rangle\approx\left(\frac{T}{T_c}\right)
\left(\frac{mg}{4\pi\hbar^2}\right)^{1/2}\log{N}.
\label{fftr}   
\end{equation}
In quasi2D trapped alkali gases one can expect a value  
$\sim 10^{-2}$ or larger for the small parameter $mg/2\pi\hbar^2$, 
and the number of trapped atoms $N$ ranging from $10^4$ to $10^6$. 

Then, from Eq.(\ref{fftr}) we identify two BEC regimes.
At temperatures well below $T_c$ the phase fluctuations are small, and
there is a true condensate. 
For intermediate temperatures 
$T<T_c$ the phase fluctuations are large and, as the density fluctuations
are suppressed, one has a quasicondensate (condensate 
with fluctuating phase). 

The characteristic radius of the phase fluctuations $R_{\phi}\!\!\approx
\!\lambdabar_T\exp{(\pi\hbar^2/mT)}$, following from Eq.(\ref{ff}) under the 
condition $\langle (\delta\hat\phi(\!{\bf R}\!))^2\rangle\!\!\sim\!\!1$, greatly
exceeds the healing length. Therefore, the quasicondensate has
the same Thomas-Fermi density profile as the true condensate. 
Correlation properties at distances smaller than $R_{\phi}$ 
and, in particular, local density correlators are also the same. Hence, one expects the same reduction of 
inelastic decay rates as in 3D condensates \cite{Kagan87}. 
However, the phase coherence properties of a quasicondensate are 
drastically different.  
For example, in the MIT type \cite{Ket} of
experiment on interference of two independently prepared quasicondensates 
the interference fringes will be essentially smeared out.  
 
We now calculate the mean-field interparticle 
interaction in a quasi2D Bose-condensed gas, relying on the binary
approximation.  
The coupling constant $g$ is influenced by the trapping 
field in the direction $z$ of the tight confinement. 
For a harmonic tight confinement, the motion of two atoms interacting
with each other via the potential $V(r)$ can be still 
separated into their relative and center of mass motion.
The former is governed by $V(r)$ together with the potential 
$V_H(z)=m\omega_0^2z^2/4$ originating from the tight harmonic confinement. 
Then, similarly to the 3D case (see, e.g. \cite{AGD}), 
to zero order in perturbation theory the coupling constant is equal to
the vertex of interparticle interaction in vacuum at zero momenta
and frequency $E=2\mu$. For low $E>0$ this vertex coincides with
the amplitude of scattering at energy $E$ and, hence, is given by 
\cite{LLQ} 
\begin{equation}    \label{g}
g=f(E)=\int d{\bf r}\psi({\bf r})V(r)\psi_f^*({\bf r}).
\end{equation}
The wavefunction of the relative motion of a pair
of atoms, $\psi({\bf r})$, satisfies the Schr\"odinger equation
\begin{equation}     \label{Schr}
\left[-\frac{\hbar^2}{m}\Delta+V({\bf r})+V_H(z)
-\frac{\hbar\omega_0}{2}\right]\psi({\bf r})=E\psi({\bf r}).
\end{equation}
The wavefunction of the free $x,y$ motion $\psi_f({\bf r})=
\varphi_0(z)\exp(i{\bf q}$\mbox{\boldmath$\rho$}), 
with $\varphi_0(z)$ being
the ground state wavefunction for the potential $V_H(z)$,
\mbox{\boldmath$\rho$}$=\{x,y\}$, and $q=(2mE/\hbar^2)^{1/2}$.
As the vertex of interaction is an analytical function
of $E$, the coupling constant for $\mu<0$ is obtained by analytical
continuation of $f(E)$ to $E<0$.

The possibility to omit higher orders in perturbation theory
requires the above criterion $(mg/2\pi\hbar^2)\ll 1$.

The solution of the quasi2D scattering problem from Eq.(\ref{Schr})
contains two distance scales: the extension 
of $\psi({\bf r})$ in the $z$ direction, $l=(\hbar/m\omega_0)^{1/2}$, 
and the characteristic
radius $R_e$ of the potential $V(r)$. In alkalis 
it ranges from $20$ \AA$\,$ for Li to $100$ \AA $\,$ for Cs.
At low energies ($qR_e\ll 1$) the amplitude $f(E)$ is determined 
by the scattering of the $s$-wave for the motion in the $x,y$ plane.

We first consider the limiting case $l\!\gg\!R_e$. Then the relative motion 
of atoms in the region of interatomic interaction is not
influenced by the tight confinement, and $\psi({\bf r})$ in Eq.(\ref{g})
differs only by a normalization coefficient from the 3D wavefunction:
\begin{equation}    \label{3D}
\psi(r)=\eta\varphi_0(0)\psi_{3D}(r).
\end{equation}
In the interval, where  $R_e\!\ll\!r\!\ll l$, Eq.(\ref{3D}) takes the 
form $\psi\!=\!\psi_{{\rm as}}(r)\!=\!\eta\varphi_0(0)(1-a/r)$. 
This expression 
serves as a boundary condition at $r\rightarrow 0$ for the solution of 
Eq.(\ref{Schr}) with $V(r)=0$ ($r\gg R_e$). The latter can be expressed
through the Green function $G({\bf r},{\bf r}')$ of this equation: 
\begin{equation}    \label{Green}
\psi({\bf r})=\varphi_0(z)\exp(i{\bf q}{\mbox{\boldmath$\rho$}})
+AG({\bf r},0),
\end{equation}
The coefficients $A$ and $\eta$ are obtained 
by matching the solution (\ref{Green}) at $r\rightarrow 0$ with 
$\psi_{{\rm as}}(r)$. 

Similarly to the case of a purely
1D harmonic oscillator (see, e.g., \cite{F}), we have 
\begin{eqnarray}    \nonumber
\!G({\bf r},0)\!=\!\frac{1}{l}\int_0^{\infty}\!\!\!\!\!
dt\frac{\exp\{i(z^2\cot t/4l^2\!-\!q^2l^2t
\!-\!t/2\!+\!\rho^2/4tl^2)\}}{t\sqrt{(4\pi i)^3\sin t}}.
\end{eqnarray}
Under the condition $ql\!\ll\!1$ ($\mu\!\ll\!\hbar\omega_0$) at 
$r\!\ll\!l$ we obtain
\begin{equation}       \label{Gsmall}
G\approx \frac{1}{4\pi r}+\frac{1}{2(2\pi)^{3/2}l}\left[\log{\left(
\frac{1}{\pi q^2l^2}\right)}+i\pi\right].
\end{equation}
Omitting the imaginary part of $G$ (\ref{Gsmall}) and
comparing Eq.(\ref{Green}) with $\psi_{{\rm as}}$,
we immediately find
\begin{equation}     \label{eta}
\eta=-\frac{A}{4\pi a\varphi_0(0)}=\left[1+\frac{a}{\sqrt{2\pi} l}
\log{\left(\frac{1}{\pi q^2l^2}
\right)}\right]^{-1}.
\end{equation} 
In Eq.(\ref{g}) one can put $\psi_f=\varphi_0(0)=(1/2\pi l^2)^{1/4}$.
Then, using the well-known result $\int d{\bf r}\psi_{3D}(r)V(r)=4\pi\hbar^2a/m$,
Eqs.~(\ref{g}), (\ref{3D}) and (\ref{eta}) lead to the coupling constant 
\begin{equation}    \label{gfinal}
g=\frac{2\sqrt{2\pi}\hbar^2}{m}
\frac{1}{l/a+(1/\sqrt{2\pi})\log{(1/\pi q^2l^2)}}.
\end{equation}
For $\mu<0$, analytical continuation of Eqs.~(\ref{Gsmall}) and (\ref{gfinal})
to $E=\hbar^2q^2/m<0$ leads to the replacement $E\rightarrow |E|=2|\mu|$ 
in the definition of $q$.

The coupling constant in quasi2D depends on 
$q\!=\!(2m|\mu|/\hbar^2)^{1/2}$ and, hence, on the condensate density.
In the limit $l\!\gg\!|a|$ the logarithmic 
term in Eq.(\ref{gfinal}) is not important, and $g$ becomes density
independent. In this case the quasi2D gas can be treated as a 3D condensate with the 
density profile $\propto\exp{(-z^2/l^2)}$ in the $z$ direction.

As follows from Eq.(\ref{gfinal}), for repulsive mean-field
interaction in 3D ($a>0$) the interaction in quasi2D is also repulsive. 
For $a<0$ the dependence $g(l)$ has a resonance character
(cf. Fig.1): The coupling constant changes sign from negative
(attraction) at very large $l$ to positive for 
$l<l_*=(|a|/\sqrt{2\pi})\log{(1/\pi q^2l^2)}$. This should describe 
the case of Cs, where $a\alt -600$ \AA \cite{Csa,Chu} and
the condition $l\gg R_e$ assumed in Eq.(\ref{gfinal}) is satisfied
at $l<l_*$.  
Near the resonance point $l_*$ the
quantity $(m|g|/2\pi\hbar^2)$ becomes large, which violates
the perturbation theory for a Bose-condensed gas and makes 
Eqs.~(\ref{g}) and (\ref{gfinal}) invalid.  

For $l\alt R_e$ (except for very small $l$) we used directly 
Eqs.~(\ref{g}) and (\ref{Schr}) and calculated numerically the 
coupling constant $g$ for Li, Na, Rb, and Cs. 
The potential $V(r)$ was modeled by the Van der Waals tail, with a
hard core at a distance $R_0\ll R_e$ selected to support many bound 
states and reproduce the scattering length $a$.
The numerical results 
differ slightly from the predictions of Eq.(\ref{gfinal}). 
For Rb and Cs both are presented in Fig.1.

\vspace{-.1cm}
\begin{figure}
\epsfxsize=\hsize
\epsfbox{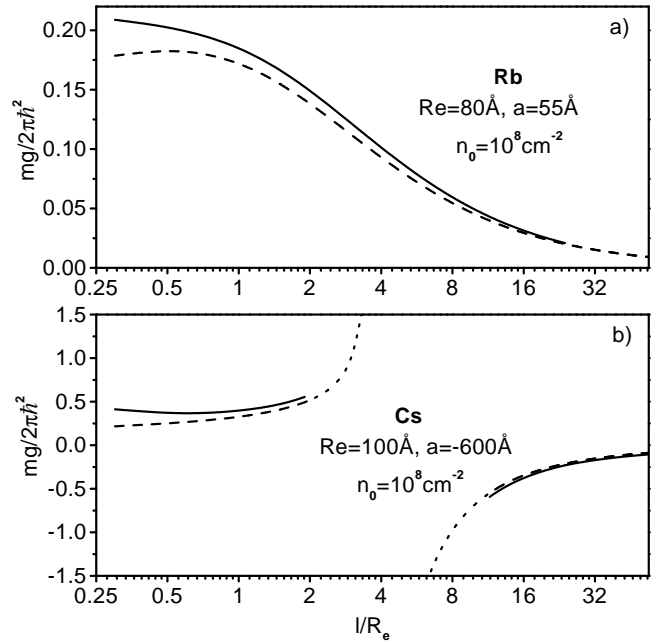}
\vspace{-3.8cm}
\caption{\protect
The parameter $mg/2\pi\hbar^2$ versus $l/R_{e}$ 
at fixed $n_{0}$ for Rb
(a) and Cs (b). Solid curves correspond to the numerical results, dashed curves to 
Eq.(\ref{gfinal}). The dotted curve in (b) shows the result of
Eq.(\ref{gfinal}) in the region where $m|g|/2\pi\hbar^2\sim 1$.  
}
\label{1}
\end{figure}

The nature of the $g(l)$ dependence in quasi2D can be understood
just relying on the values of $g$ in the purely 2D and 3D cases.
In 2D at low energies the mean-field 
interaction is always repulsive. This striking difference 
from the 3D case can be found from the solution of the 2D scattering 
problem in \cite{LLQ} and originates from the 2D kinematics:
At distances, where $R_e\ll\rho\ll q^{-1}$ ($q\rightarrow 0$),
the solution of the Schr\"odinger equation for the (free) relative 
motion in a pair of atoms  reads 
$\psi\propto\log(\rho/d)/\log(1/qd)$ ($d>0$).
We always have $|\psi|^2$ increasing with $\rho$, 
unless we touch resonances corresponding to the presence of a bound 
state with zero energy ($d\rightarrow\infty$). This means 
that it is favorable for particles to be at larger $\rho$, i.e. 
they repel each other.  

In quasi2D for very large $l$ the sign of the interparticle interaction 
is the same
as in 3D. With decreasing $l$, the 2D features in the
relative motion of atoms become pronounced, which is described by the
logarithmic term in Eq.(\ref{gfinal}). Hence, for $a>0$ the interaction
remains repulsive, whereas for $a<0$ the attraction turns to repulsion.  

The obtained results are promising for tunable BEC in quasi2D gases,
based on variations of the tight confinement and, hence, $l$.
However, as in the MIT studies of tunable 3D BEC by using Feshbach
resonances \cite{KetF}, an "underwater stone" concerns inelastic
losses: Variation of $l$ can change the rates of inelastic processes. 
For optically trapped atoms in the lowest Zeeman state the most important decay 
process is 3-body recombination. 

This process occurs at interparticle distances 
$r\alt {\rm max}\{R_e,|a|\}$ \cite{FRS,Greene}.
We will restrict ourselves to the case where $l\agt |a|$ and
is also significantly larger than $R_e$. Then the character of recombination
collisions remains 3-dimensional, and one can treat them 
in a similar way as in a 3D gas with the density profile 
$(n_0/\sqrt{\pi}l)\exp{(-z^2/l^2)}$.

However, the normalization coefficient of the 
wavefunction in the incoming channel will be influenced by the tight
confinement. Relying on the Jastrow approximation, we write this
wavefunction as a product of the three wavefunctions $\psi({\bf r}_{ik})$,
each of them being a solution of the binary collision problem 
Eq.(\ref{Schr}). In our limiting case the solution is given by Eq.(\ref{3D})
divided by $\varphi_0(0)$ to reconstruct the density profile in the $z$
direction. 
The outgoing wavefunction remains the same as in 3D,
since one has a molecule
and an atom with very large kinetic energies. 

Thus, in the Jastrow approach we have an additional factor $\eta^3$
for the amplitude and $\eta^6$ for the probability of recombination in
a quasi2D gas compared to the 3D case. Averaging over the density profile 
in the $z$ direction, we can relate the quasi2D rate constant $\alpha$ to the 
rate constant in 3D (see \cite{Greene} for a table of $\alpha_{3D}$ in alkalis):
\begin{equation}    \label{alpha}
\alpha=(\eta^6/\pi l^2)\alpha_{3D}.
\end{equation}
As $\eta$ is given by Eq.(\ref{eta}), for $a>0$ 
the dependence $\alpha(l)$ is smooth. 
For $a\!<\!0$ the rate constant $\alpha$ peaks at 
$l\!\approx\!l_*$
and decreases as $l^4/(l_*-l)^6$ at smaller $l$.
This indicates a possibility to reduce recombination
losses while maintaining a repulsive mean-field interaction ($g\!>\!0$).
For Cs already at $l\!\approx\!200$ \AA $\,$ ($l_*\!\approx\!500$\AA) 
we have $\alpha\!\sim\!10^{-17}$ cm$^4$/s, and at densities 
$10^8$ cm$^{-2}$ the life-time $\tau\!>\!1$ s.

The predicted possibility to
modify the mean-field interaction and reduce 
inelastic losses by varying the frequency of the tight confinement
opens new handles on tunable BEC in quasi2D 
gases. These experiments can be combined with measurements of 
non-trivial 
phase coherence properties of condensates with fluctuating phase.      
  
We acknowledge fruitful discussions with C. Salomon, I. Bouchoule, and
J.T.M. Walraven.
This work was financially supported by the Stichting voor Fundamenteel 
Onderzoek der Materie (FOM), by the CNRS, by INTAS, and by the 
Russian Foundation for Basic Studies. 

$^{*}$ L.K.B. is an unit\'e de recherche de l'Ecole Normale Sup\'erieure
et de l'Universit\'e Pierre et Marie Curie, associ\'ee au CNRS.

\end{document}